\documentclass[twocolumn,showpacs,amsmath,pra]{revtex4-1}

\usepackage{graphicx}
\usepackage{subfigure}
\usepackage{epsfig}
\usepackage{dcolumn}
\usepackage{bm}
\usepackage{color}
\usepackage{xcolor}

\usepackage{amsmath, amssymb, dsfont}

\def\be{\begin{equation}}       \def\ee{\end{equation}}
\def\bea{\begin{eqnarray}}      \def\eea{\end{eqnarray}}
\def\ba{\begin{array}}
\def\ea{\end{array}}
\def\bnum{\begin{enumerate} }
\def\enum{\end{enumerate}}

\def\=>{\Rightarrow}
\def\>{\rightarrow}

\def\eye2{Fathbb{I}}

\renewcommand{\>}{\rangle}

\newcommand{\al}{\alpha}

\begin{document}

\title{Topological photonic crystal with equifrequency Weyl points}
\author{Luyang Wang$^{1}$, Shao-Kai Jian$^{1}$, Hong Yao$^{1,2}$}
\affiliation{$^1$Institute for Advanced Study, Tsinghua University, Beijing 100084, China\\
$^2$Collaborative Innovation Center of Quantum Matter, Beijing 100084, China
}
\date{\today}

\begin{abstract}
  Weyl points in three-dimensional photonic crystals behave as monopoles of Berry flux in momentum space. Here, based on general symmetry analysis, we show that a minimal number of four symmetry-related (consequently equifrequency) Weyl points can be realized in time-reversal invariant photonic crystals. We further propose an experimentally-feasible way to modify double-gyroid photonic crystals to realize four equifrequency Weyl points, which is explicitly confirmed by our first-principle photonic band-structure calculations. Remarkably, photonic crystals with equifrequency Weyl points are qualitatively advantageous in applications including angular selectivity, frequency selectivity, invisibility cloaking, and three dimensional imaging.
\end{abstract}
\maketitle
	
Eighty six years after Hermann Weyl derived the famous equation named after him\cite{weyl-1929}, Weyl points, near which (quasi)particles are described by that equation, have been experimentally  observed in condensed matter systems\cite{lv-2015a,xu-2015a,yang-2015,lv-2015b,xu-2015b} and photonic crystals (PhC)\cite{lu-2015}. Weyl points are twofold degenerate points in crystal momentum space where the spectrum disperses linearly\cite{volovik-2009,wan-2011, xu-2011,burkov-2011,yang-2011, halasz-2012,zhang-2014, liu-2014,weng-2015, huang-2015,hirayama-2015}. They act like monopoles of Berry flux, and accordingly, a chirality of $\pm 1$ can be associated to each of them. Due to their topological nature, many unusual phenomena like surface arc states emerge on the boundary of such systems\cite{wan-2011}. Since Weyl points act like magnetic monopoles emitting Berry flux in crystal momentum space, they must appear in pairs in order to maintain the neutrality of the whole Brillouin zone\cite{nielsen-1983}. Because the Berry curvature in crystal momentum space is odd under the PT transformation [the product of inversion (P) and time reversal (T)], either inversion or time reversal symmetry (or both) must be broken to obtain isolated Weyl points. In general, the minimal number of Weyl points in systems breaking the time-reversal symmetry is two while in systems respecting T symmetry it is four, the latter of which was recently experimentally observed\cite{lu-2013,lu-2015}.

Photons in vacuum are described by the Maxwell equations and are linearly dispersed with two frequency-degenerate polarizations; this vacuum state of light may be considered as two overlapping Weyl points with opposite monopole charges\cite{bliokh-2015}. It is remarkable that isolated Weyl points with finite frequencies were theoretically shown in photonic crystals\cite{lu-2013}. Other interesting topological phenomena theoretically proposed in the context of photonic crystals include quantum Hall effect\cite{haldane-2008,raghu-2008, wang-2008,wang-2009}, quantum spin Hall effect\cite{he-2014a, he-2014b}, and photonic topological insulators\cite{khanikaev-2012}. Recently, it was reported that four Weyl points were experimentally observed in a photonic crystal which respects the time-reversal symmetry\cite{lu-2015}. However, these Weyl points are not symmetry-related such that they occur at different frequencies.

If all Weyl points in a photonic crystal can be related by symmetry, they must locate at the same frequency. Such equifrequency Weyl points in a photonic crystal are highly desired because of the following qualitative advantages. In the topological photonic crystal with equifrequency Weyl points, the momentum degeneracy at the Weyl point frequency is extremely reduced to discrete momentum points, similar to the case of zero-frequency degeneracy in vacuum. In other words, inside such a photonic crystal, a photon at this frequency have to propagate in discrete momenta. In addition, the linear dispersion near Weyl points gives rise to an effective refractive index, which can be either positive or negative. If equifrequency Weyl points are realized in PhC, these properties may pave new avenue to realize novel optical devices in the future.

Here, we show that a minimal number of equifrequency Weyl points can be realized in a PhC with time-reversal symmetry. Our starting point is a quadratic bands touching (QBT) \cite{sun-2009,chong-2008} in double-gyroid (DG) PhC with both inversion and time-reversal symmetries. Then, we break the inversion symmetry to split the QBT at $\Gamma$ point and obtain four symmetry-related Weyl points at $k_x$ and $k_y$ axes. They are related by the remaining $S_4$ symmetry and hence locate at the same frequency. Note that it is the minimal number of Weyl points a time-reversal invariant system can have. We further explicitly obtain these four equifrequency Weyl points by first-principle calculations of the photonic band structure. Finally, we discuss the potential applications of equifrequency Weyl points with special attentions to its discrete Weyl-points property and linear dispersions, including angular/frequency selectivity, invisibility cloaking and three dimensional imaging.

\begin{figure*}
	\centering
    	\subfigure[]{\includegraphics[width=3.8cm]{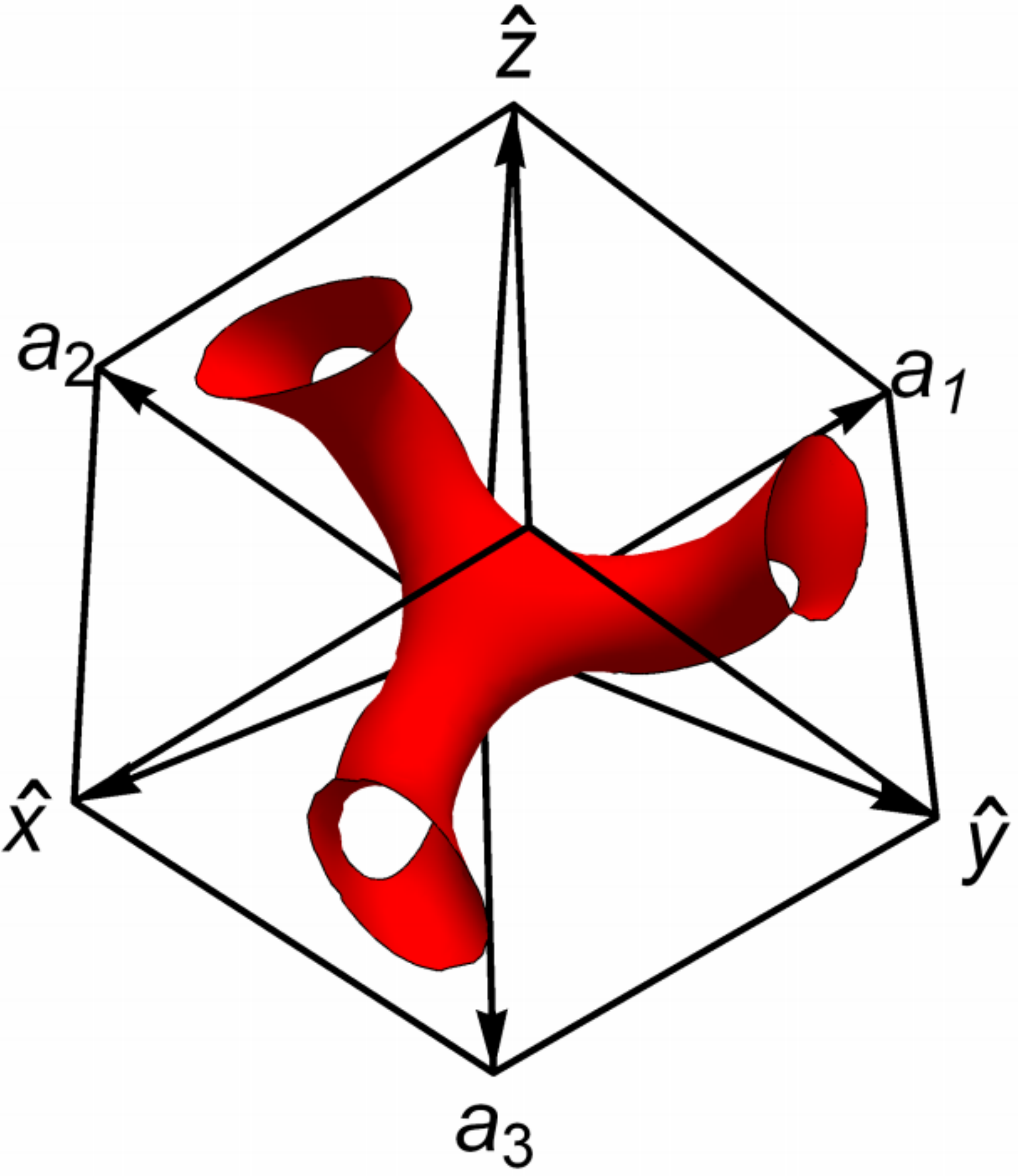}\label{SG}}~~
    	\subfigure[]{\includegraphics[width=3.8cm]{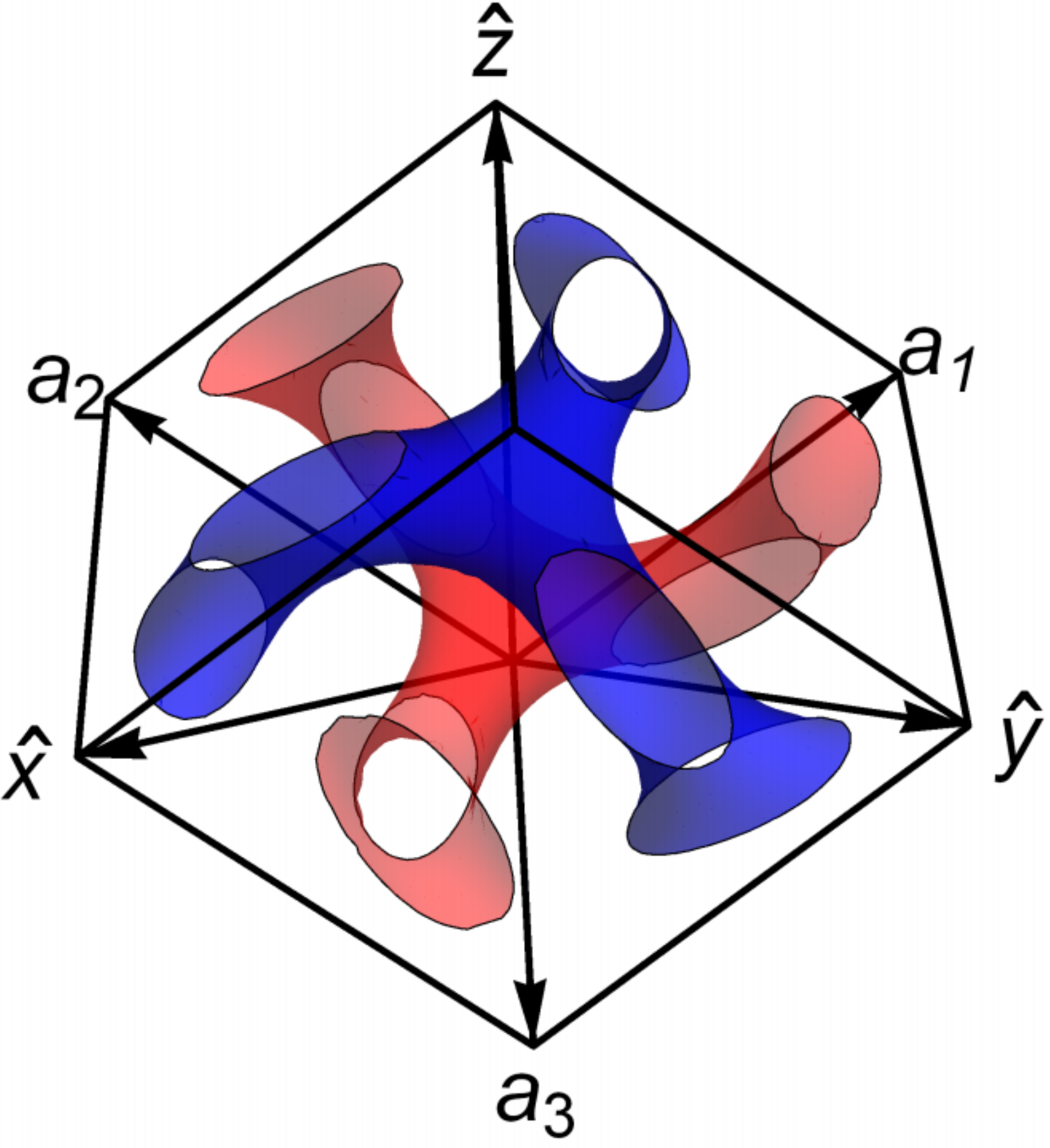}\label{DG}}~~
	\subfigure[]{\includegraphics[width=3.8cm]{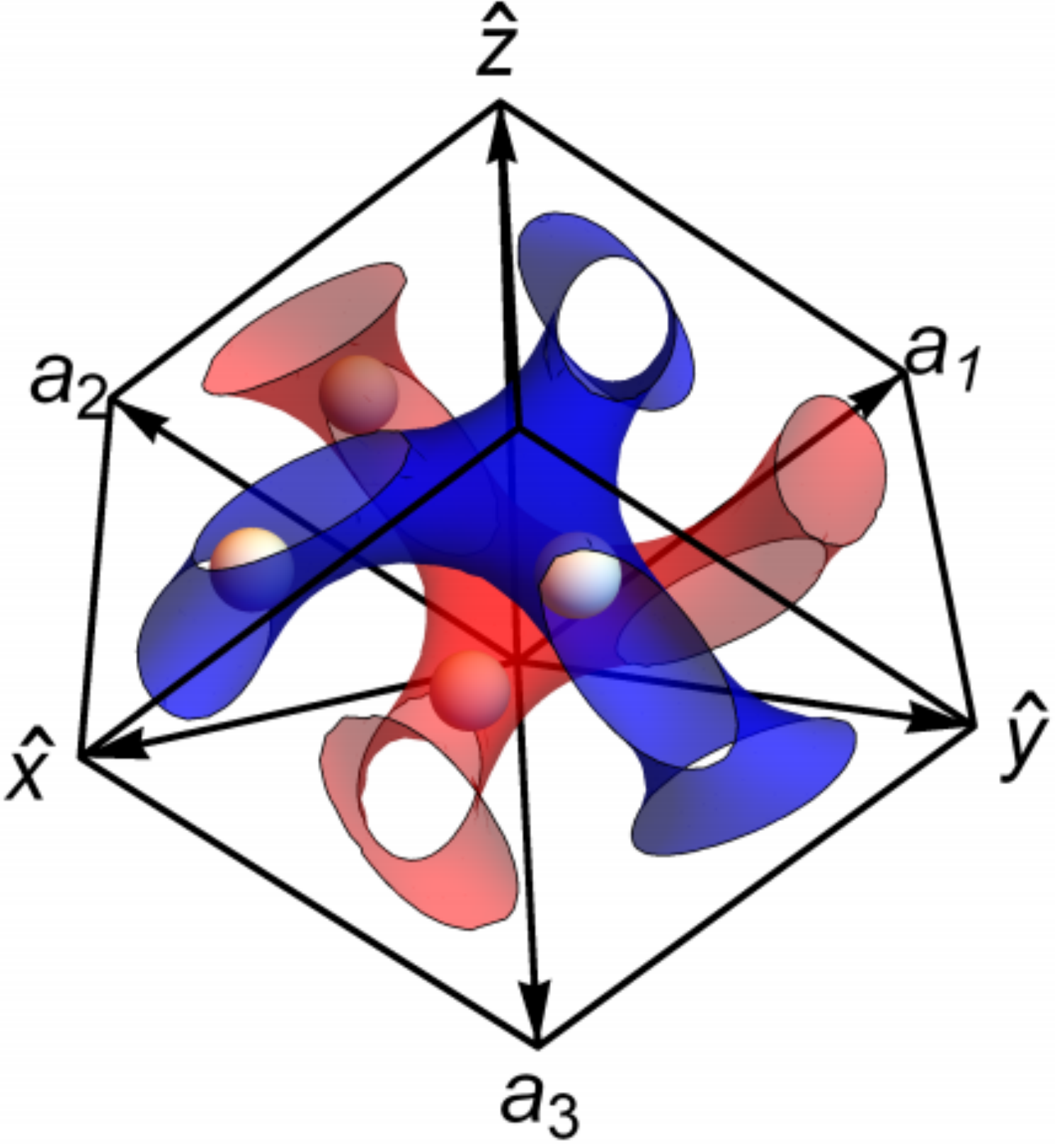}\label{DGair}}~~
	\subfigure[]{\includegraphics[width=4.4cm]{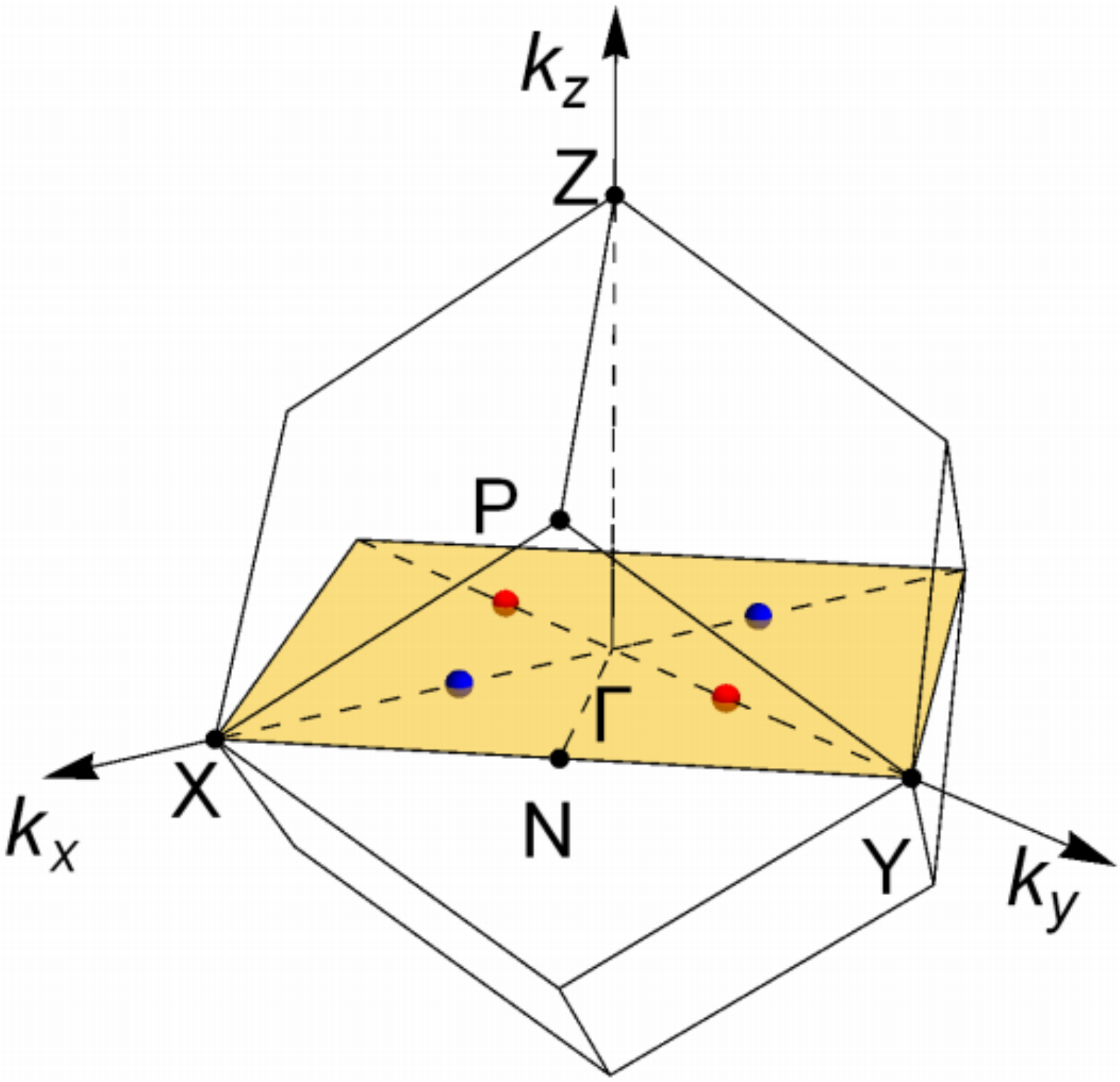}\label{BZ}}
	\caption{Gyroid photonic crystal has a body-center cubic structure. The lattice vector are given by ${\bf a}_1=(-1,1,1)\frac{a}{2},{\bf a}_2=(1,-1,1)\frac{a}{2}$, and ${\bf a}_3=(1,1,-1)\frac{a}{2}$. (a) Primitive cell of single-gyroid photonic crystals. The red region for $g(\vec r)>1.1$ is occupied by dielectric materials. (b) Primitive cell of double-gyroid photonic crystals. The red ($g(\vec r)>1.1$) and blue ($g(\vec r)<-1.1$) regions represent two single gyroids related by the inversion symmetry. (c) Air spheres modified double-gyroid structure. The positions of air spheres are given by $(\frac14, -\frac18, \frac12) a$, $(\frac14, \frac18,0) a$, $(\frac58, 0, \frac14) a$, and $(\frac38, \frac12, \frac14) a$. The radius of the air spheres is $r=0.07a$. (d) The first Brillouin zone of the bcc lattice.}
\end{figure*}

{\it Symmetry analysis}. The DG structure consists of two single gyroids (SG). A SG has a body-centered cubic (bcc) structure and is described by the function\cite{wohlgemuth-2001} $g(\vec{r})=\sin (2\pi x/a) \cos (2\pi y/a)+\sin (2\pi y/a) \cos (2\pi z/a)+ \sin (2\pi z/a) \cos (2\pi x/a)$, where $a$ is the lattice constant.  Fig. \ref{SG} shows the SG structure where dielectric material occupies the region of $g(\vec r)>g_0$ with $g_0=1.1$ (the region inside the red surface). It has $I4_1 32$ space group symmetry where P symmetry is broken. Two P symmetry-related SG make up a DG which has the space group $I \bar{a} 3d$ that is the direct product of space group $I4_1 32$ and inversion. The DG structure is shown in Fig. \ref{DG}, where red [$g(\vec r)>g_0$] and blue [$g(\vec r)<-g_0$] represent two SG related by inversion.

It was shown that the 3$^{\text{rd}}$, 4$^{\text{th}}$, and 5$^{\text{th}}$ bands touch at $\Gamma$ point\cite{maldovan-2002} in a DG PhC, forming a three-dimensional representation, and that the dispersion around the touching point is quadratic because terms linear in $k$ are absent as required by the T and P symmetry. The effective Hamiltonian near $\Gamma$ point up to quadratic level is given by\cite{luttinger-1956}
\bea
H(\vec{k}) = \al_1 |\vec{k}|^2 + \al_2 \sum_i k_i^2 L_i^2 + \al_3 \big[k_x k_y \{ L_x, L_y \} + c.p.\big], ~~~\label{h0}
\eea
where $L_i$ is a 3$\times$3 matrix of spin-1 representation (see Appendix for details), $\al_i$ is a constant characterizing the dispersion near $\Gamma$ point, $\{\}$ denotes anticommutator, and $c.p.$ means cyclic permutations.

The QBT at $\Gamma$ point is a perfect starting point to realize Weyl points. In order to achieve this, at least one of T or P symmetry should be broken. Since it is easier to break P symmetry in photonic crystals, we consider P-breaking perturbations that maintain T symmetry. In general, Weyl points are not equifrequency (symmetry-related), and they have different frequencies/energies. In the absence of P symmetry, mirror symmetry and/or $S_4$ symmetry should be maintained to relate Weyl points with different chiralities. Thus, to realize equifrequency Weyl points, we add perturbations that lower $O_h$ space group to its subgroup $D_{2d}$ for which $S_4$ and $\sigma_d$ are maintained while P symmetry is broken.

The most relevant perturbation near $\Gamma$ point is a constant term which does not depend on $\vec k$. As there is only one such term (proportional to $L_z^2$) which respects the $D_{2d}$ symmetry (see Appendix), we first consider such a perturbation: $H_1= \beta_1 (3L_z^2-2)$, where $\beta_1$ is a constant describing the strength of the perturbation. Note that the perturbation $H_1$ respects $D_{4h}$ symmetry. Due to the explicit breaking of the $O_h$ symmetry to the $D_{4h}$ one, the QBT at the $\Gamma$ point is split and a line node located at $k_z=0$ plane is generated. Note that P and T symmetry are unaffected by this perturbation, as well as the mirror symmetry $\sigma_h$.  The line node is protected by $\sigma_h$ that sends $z \rightarrow -z$. Formally, a topological invariant can be defined in the presence of P and T symmetry and explains the stability of line nodes\cite{fang-2015}.

Linear terms in $\vec k$ are considered as next relevant perturbations. Since $\vec k$ is odd under T operation, the only suitable matrices are $L_i$ which are also odd under T operation. Even though there are nine possible $k_iL_j$ terms, restricting to the $D_{2d}$ symmetry gives rise to only one such term given by $H_2(\vec{k})=\beta_2(k_x L_x- k_y L_y)$, where $\beta_2$ is a constant capturing the strength of the perturbation. This perturbation breaks both P and $\sigma_h$ symmetries. As a consequence, the line node splits except in four discrete points, which are located on the $k_x$ and $k_y$ axis. The remaining four touching points on the $k_x$ and $k_y$ axis are stable because they are protected by the $C_{2T}\equiv C_2\cdot T$ symmetries of the $k_x$ or $k_y$ axis, where $C_2$ represent the $\pi$ rotation along the $k_x$ or $k_y$ axis \cite{ruan-2015,ruan-2016}.

\begin{figure}[t]
	\centering
    	\subfigure[]{\includegraphics[width=2.3cm]{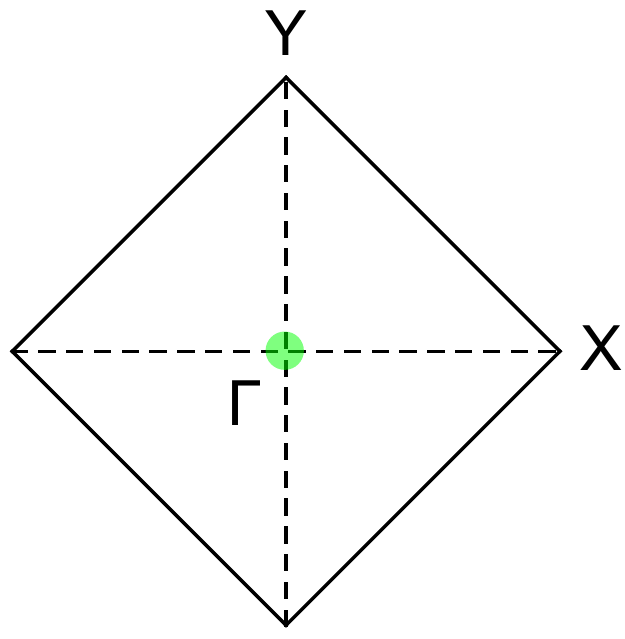}\label{qbt}}
	\subfigure[]{\includegraphics[width=6cm]{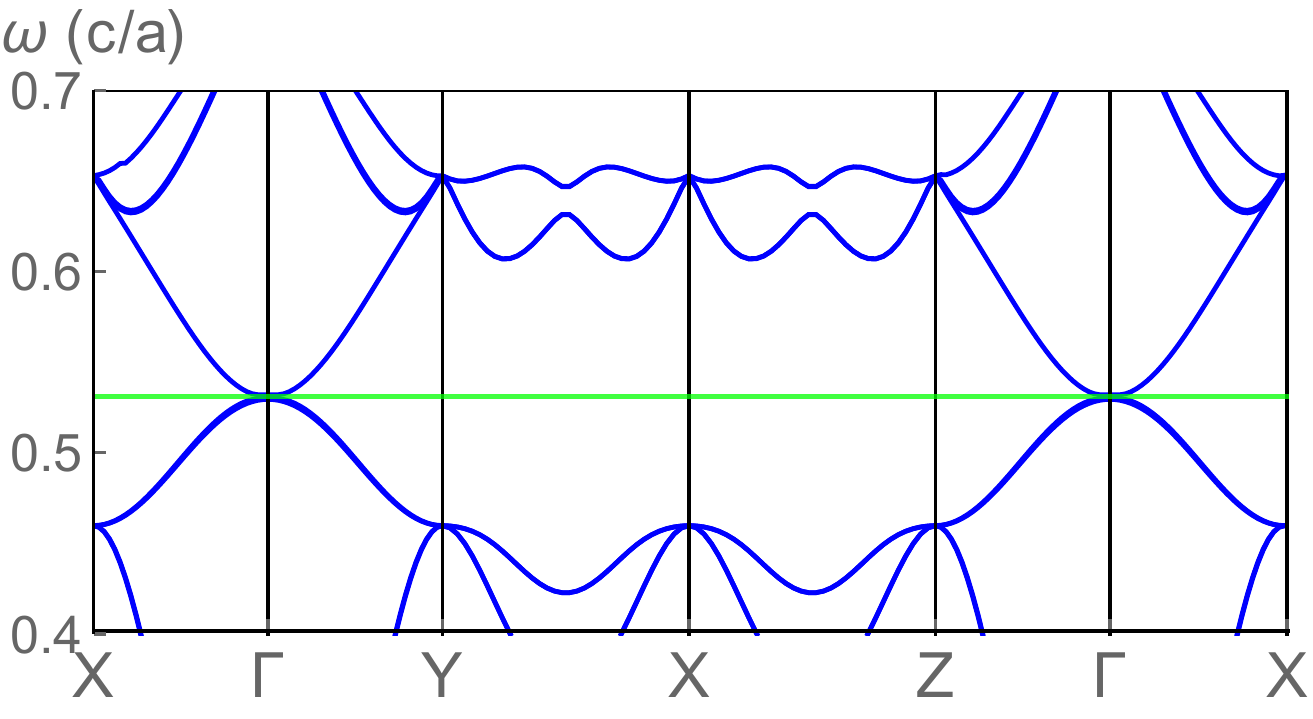}\label{band0}}
	\subfigure[]{\includegraphics[width=2.3cm]{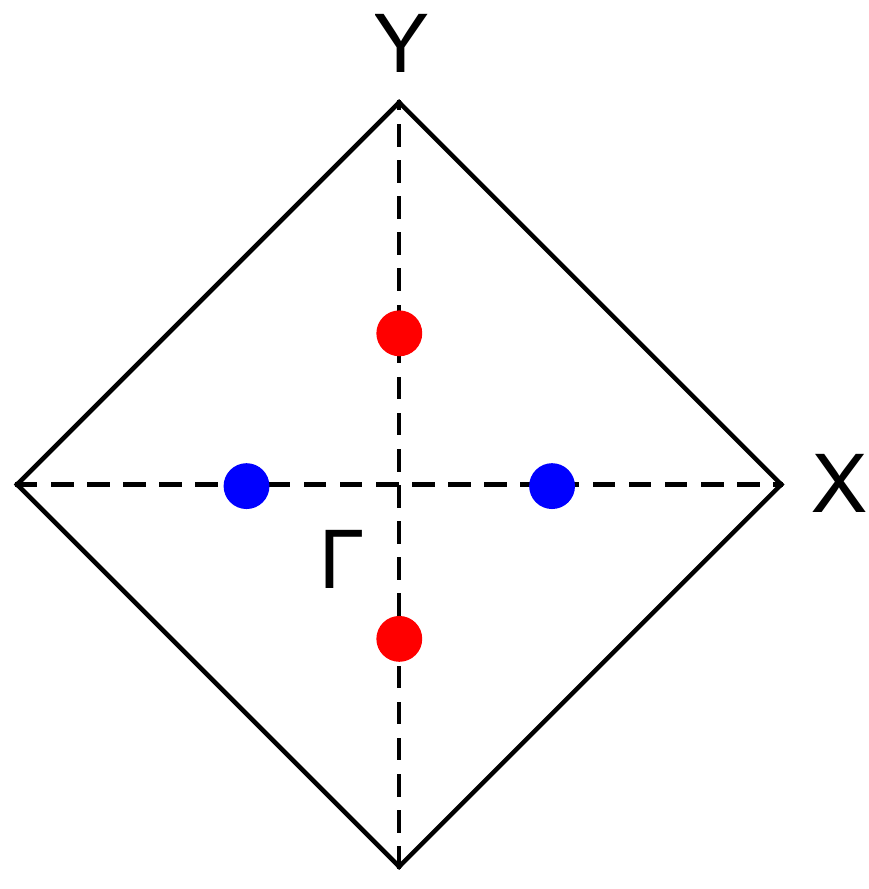}\label{weyl}}
    	\subfigure[]{\includegraphics[width=6cm]{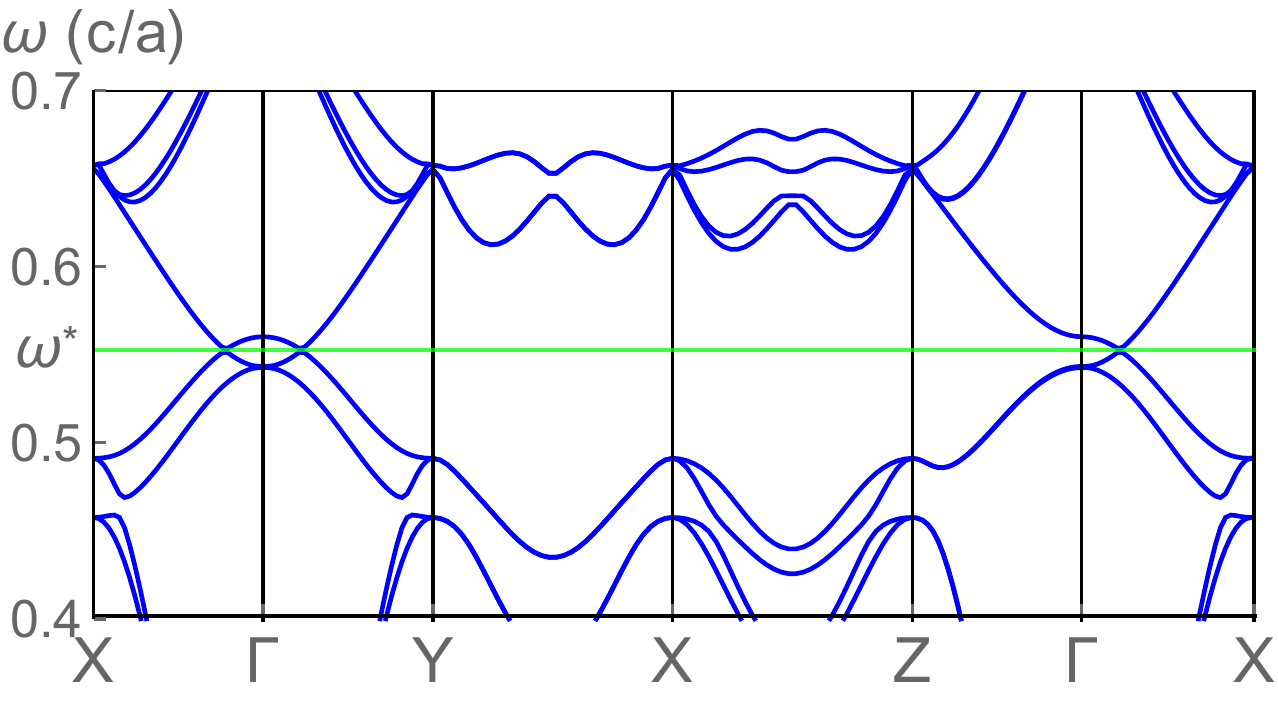}\label{band4}}
	\caption{Calculated first eight bands of DG PhC with or without air spheres. (a) The QBT is shown in the $k_z=0$ plane by a green spot. (b) The band dispersions of DG PhC without air spheres. The dispersions are highly symmetric owing to the $O_h$ symmetry and the QBT at $\Gamma$ point is manifest. (c) Schematic representation of equifrequency Weyl points located at the $k_x$ and $k_y$ axes. They can be understood as the splitting of the QBT by breaking the inversion symmetry. (d) The calculated band dispersion of modified DG PhC. Since the $O_h$ is broken to $D_{2d}$, the QBT at $\Gamma$ point splits into the four equifrequency Weyl points, located at X-$\Gamma$ and $\Gamma$-Y.}
\end{figure}

The effective Hamiltonian around one of the aforementioned gapless points, e.g., $(k_x^\ast,0,0)$, is given by the Weyl Hamiltonian(see Appendix for details)
\bea
H_{\text{Weyl}}(\vec{k}) = d_0(\vec k)\sigma^0 + v_x  k_x \sigma^z+ v_y  k_y \sigma^y+ v_z  k_z \sigma^x,
\eea
where the velocity $v_i$'s are determined by $\al_i$ and $\beta_i$, $\vec k$ is the momentum away from the Weyl point, and $d_0(\vec k)= v'_x k_x$.  The other three Weyl points are related to this one by the $S_4(z)$ symmetry so that all Weyl points are symmetry-related. In the perturbative region as we calculate, the Weyl points are present only when $3 \beta_1 \al_2+ \beta_2^2 >0$. Note that the Weyl points are type-I (type-II) when $|v_x'|<|v_x|$ ($|v_x'|>|v_x|$) \cite{soluyanov-2015}. The type of Weyl points generally depends on $\delta=\frac{\al_2}{\al_1}$, $\beta_1$, and $\beta_2$. For $\delta<-2$ ($\delta>-1$), the Weyl points are always only type-I (type-II), independent on $\beta_1$ and $\beta_2$; for $-2<\delta<-1$, the type of Weyl points is further depends on the $\beta_1$ and $\beta_2$ (see Appendix for details).

{\it Equifrequency Weyl points in modified double-gyroid}. As analyzed above, four equifrequency Weyl points can occur by lowering the $O_h$ space group of DG structure to the $D_{2d}$ group. Here, we propose a such modification of DG structure in order to realize equifrequency Weyl points. As shown in Fig. \ref{DGair}, we introduce four air spheres in a unit cell of DG PhC; one of them is located at $(\frac{1}{4},\frac{7}{8},\frac{1}{2})a$, and the other three Weyl points are related to it by $S_4(z)$ transformation. In the presence of air spheres, the symmetry is lowered to $D_{2d}$.

To show that Weyl points are present in such modified structures, we implement numerical band calculations\cite{mpb}. In the calculations, the dielectric material occupies the region where $|g(\vec r)|>1.1$, while the other region is occupied by air. The dielectric constant of air is set to be $\epsilon_{\text{air}}=1.0$, and of gyroid $\epsilon_{\text{gyr}}=13.0$. We calculate the eight lowest bands in the DG PhC with and without air spheres. The first Brillouin zone (BZ) of bcc lattice is shown in Fig. \ref{BZ}. The eight lowest bands along X-$\Gamma$-Y-X-Z-$\Gamma$-X are calculated. The band dispersion in the DG PhC without modification is shown in Fig. \ref{band0}, where the band frequency is expressed in unit of $c/a$ with $c$ the speed of light in vacuum. The QBT around the $\Gamma$ point is manifest as clearly shown in Fig. \ref{band0}. Due to the presence of $O_h$ symmetry, X, Y and Z points in the BZ are equivalent, and the dispersion is symmetric. From the calculated data, the parameters in Eq. ($\ref{h0}$) can be fitted: $\al_1\approx 4.5 \times 10^{-3}$, $\al_2\approx -8.0 \times 10^{-3}$, and $\al_3\approx -1.2\times 10^{-2}$ (in units of $ac$) and accordingly, $\delta=\frac{\alpha_2}{\alpha_1}\approx -1.8$. Thus, $-2<\delta<-1$, both type-I Weyl points and type-II are possible. However, in the specific parameter range calculated by us, only type-I Weyl points occur (see below).

Fig. \ref{band4} shows the band dispersion in the modified DG PhC. The radius of air spheres is set to be $r/a=0.07$. We can see from the figure that the QBT in $\Gamma$ splits due to the lowing of the lattice symmetry from the $O_h$ to $D_{2d}$ group. The crossings of two bands from X-$\Gamma$ and $\Gamma$-Y in the shaded region are the desired equifrequency Weyl points, which are related by the $S_4$ symmetry. Fig. \ref{weyl} schematically shows the positions of four equifrequency Weyl points at $\omega^*$. In a photonic crystal, both the frequency and the crystal momentum scale with $1/a$, which makes the band structure applicable to a large range from microwave to optical frequency. To adjust the frequency and momentum independently, one can first adjust $a$ and then the size of the air spheres -- the larger the spheres are, the further the Weyl points are from $\Gamma$. For instance, we obtain $\omega^*\approx 1.66\times10^{11}$Hz for $a=1.0$mm and $r/a=0.07$. In this case, the range of linear dispersion is estimated a few GHz, but it can be made larger by increasing the radius of the air spheres.

\begin{figure}
\begin{minipage}{.24\textwidth}
\subfigure[]{\includegraphics[width=3.4cm]{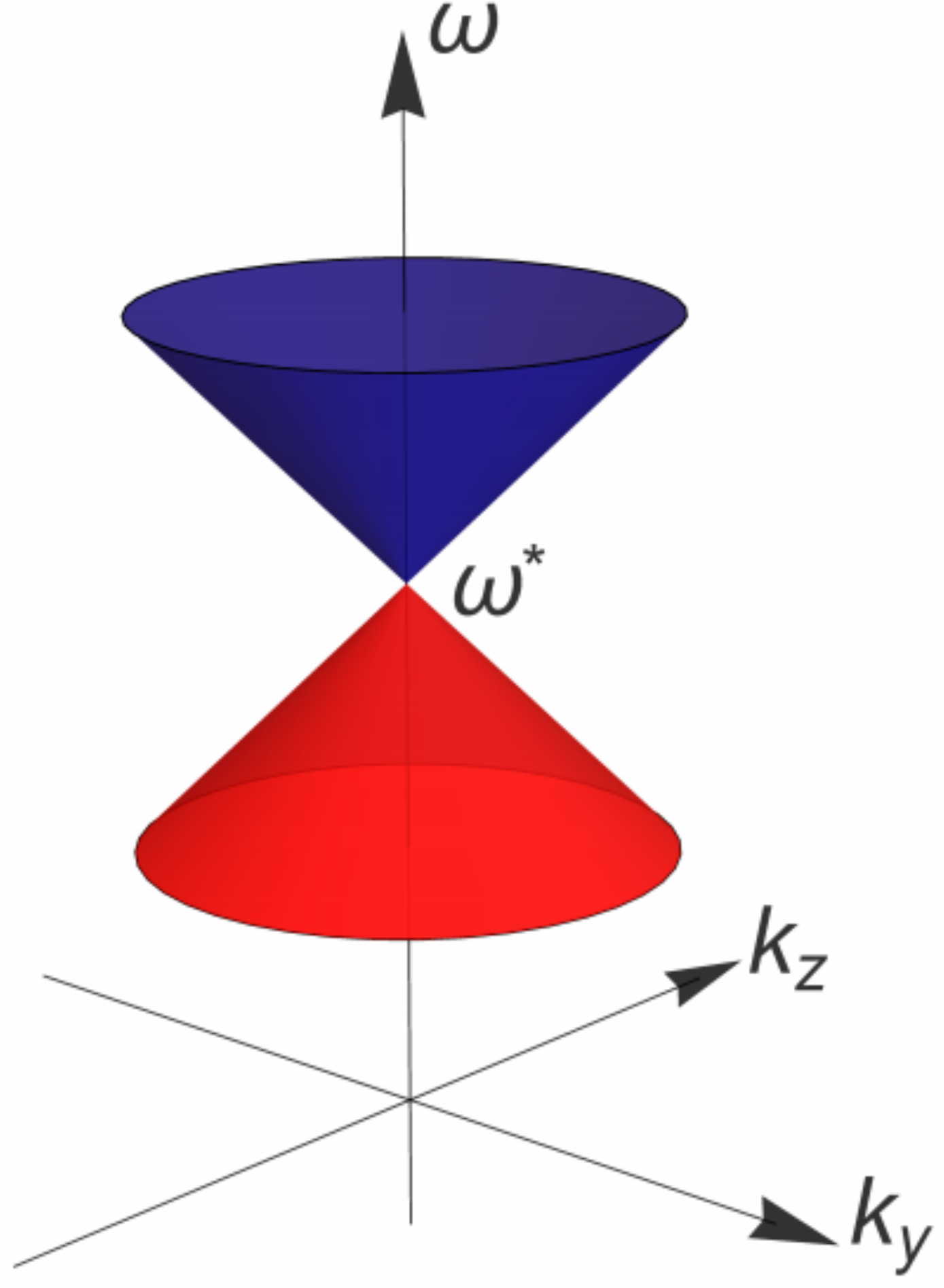}\label{cone}}\\[1ex]
\subfigure[]{\includegraphics[width=2.4cm]{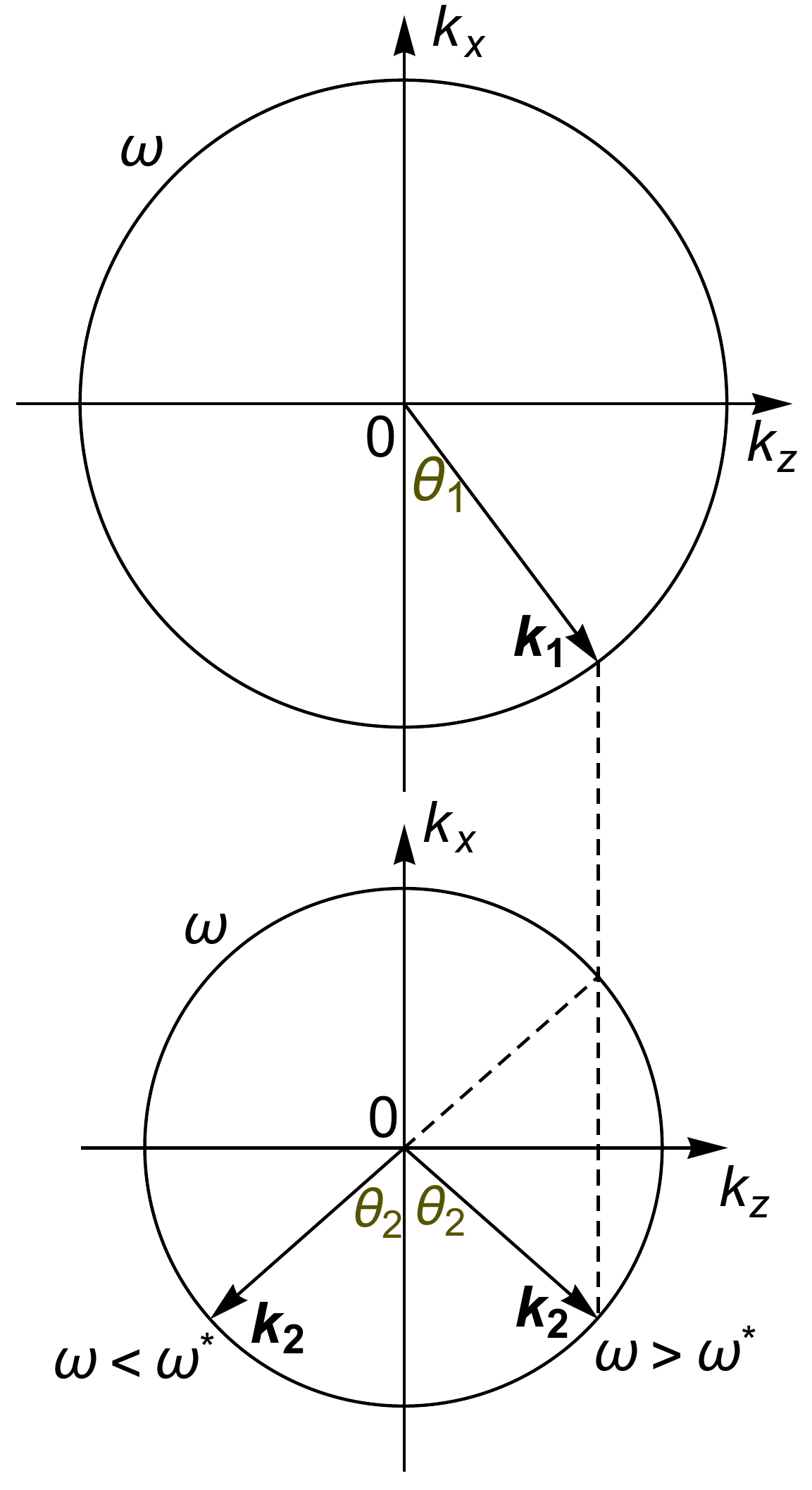}\label{noR}}
\end{minipage}%
\begin{minipage}{.24\textwidth}
\subfigure[]{\includegraphics[width=3.7cm]{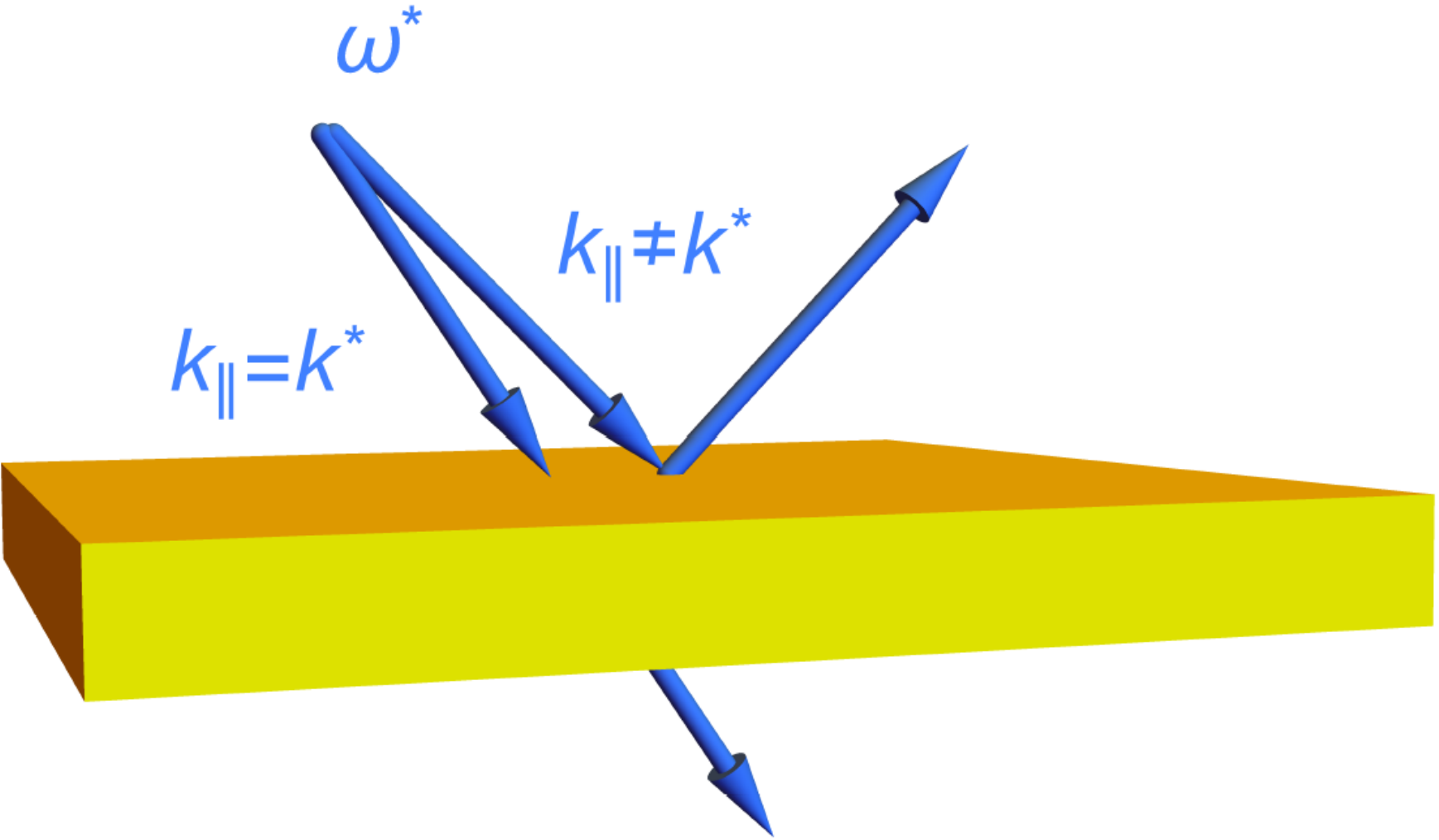}\label{angular1}}\\[5ex]
\subfigure[]{\includegraphics[width=3.9cm]{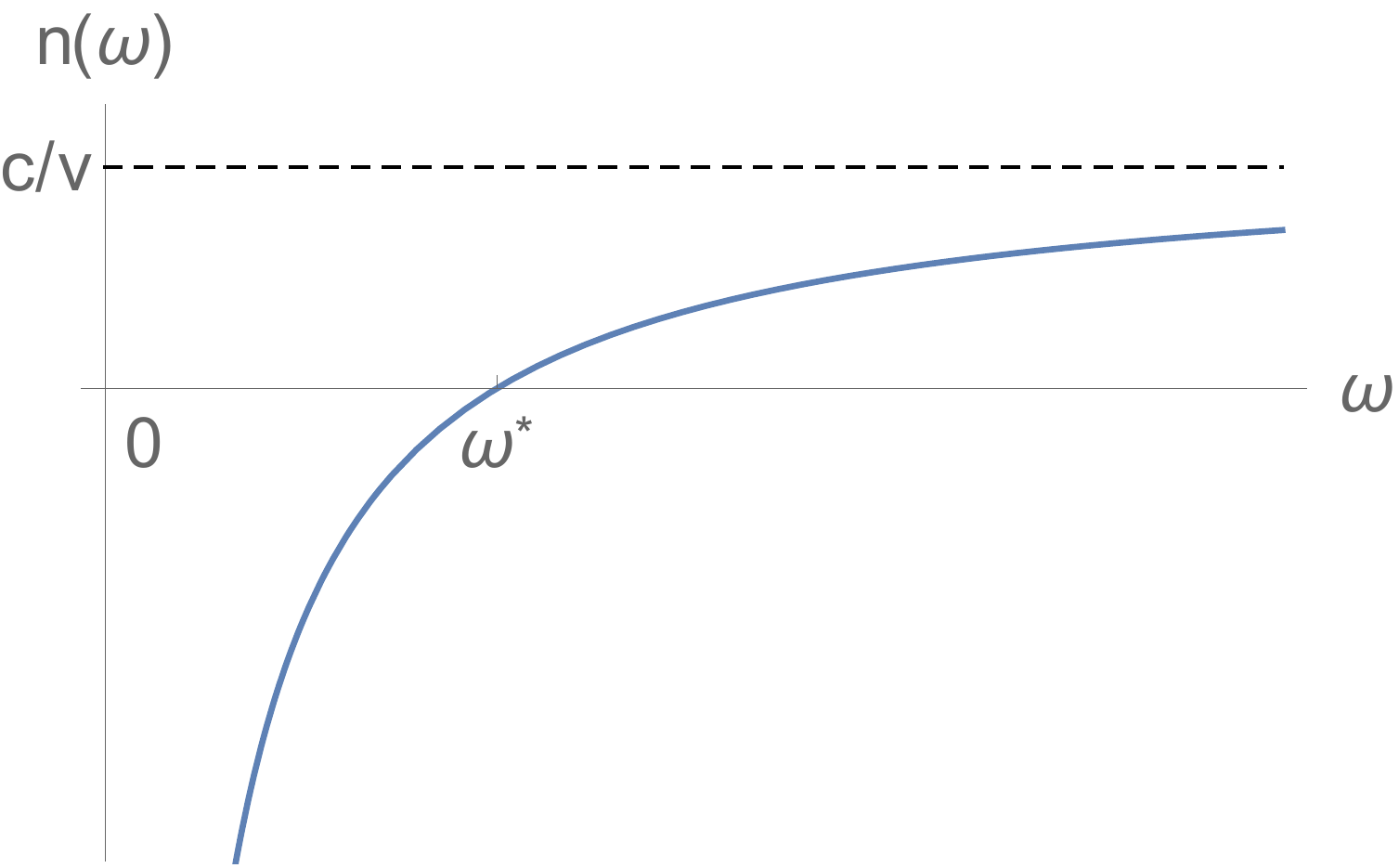}\label{r}}\\[4ex]
\subfigure[]{\includegraphics[width=3.7cm]{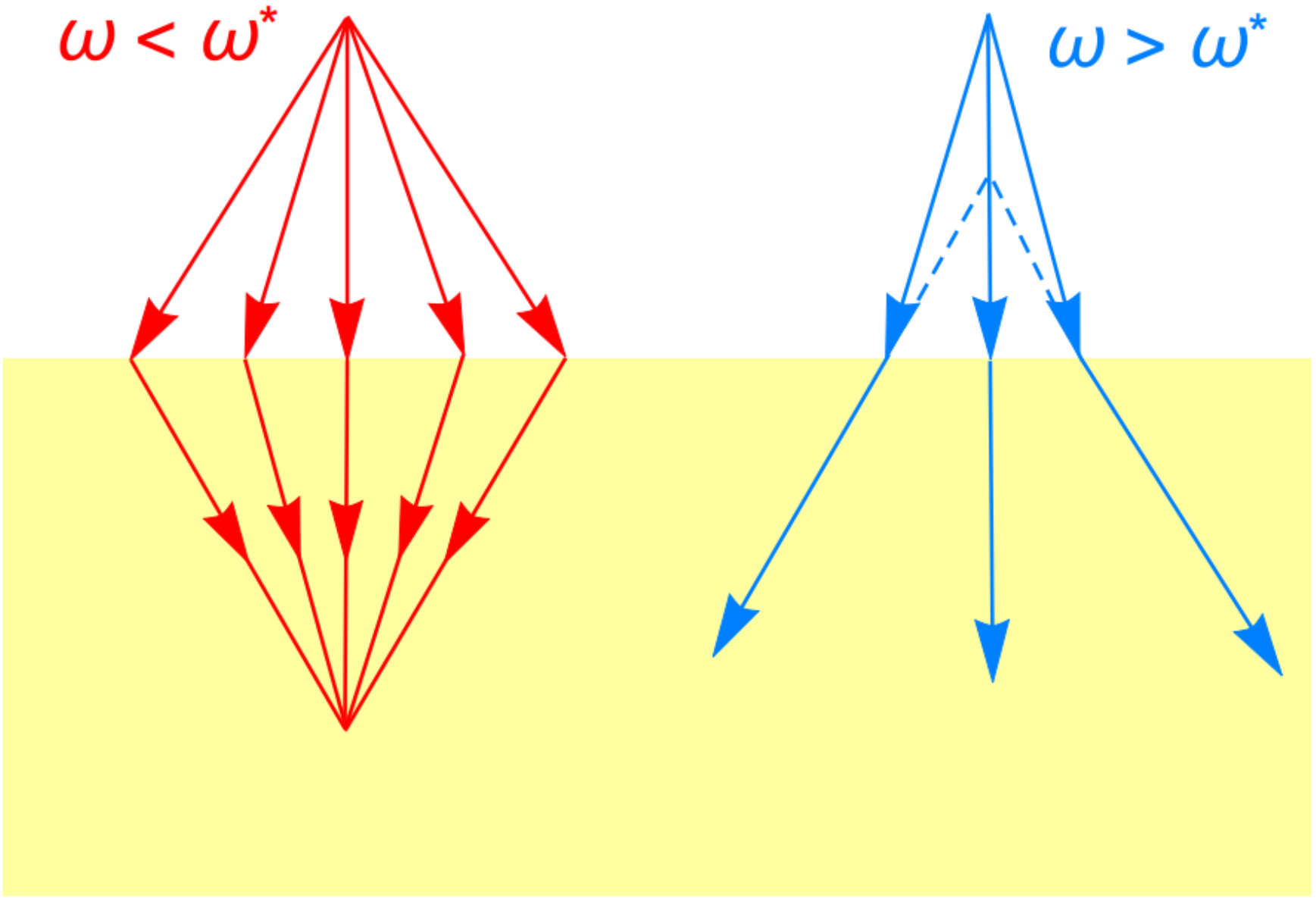}\label{images}}
\end{minipage}
\caption{(a) The band structure near one Weyl point. The upper and lower cone have a positive and negative refractive index, respectively. (b) The $k_y=0$ cross-section of equifrequency surfaces in vacuum (upper circle) and in the photonic crystal (lower circle)\cite{notomi-2000}, if $\omega\neq\omega^*$. When the light shines from $\hat{x}$ direction and is confined in $xz$ plane, the $k_z$ component is conserved at the boundary. Here, $\theta_1$ and $\theta_2$ are the incident angle and the refractive angle, respectively. (c) At $\omega^*$, only the rays at particular directions can transmit, while those at other directions are totally reflected. (d) If the Weyl cone is nearly isotropic, the refractive index is well-defined and can be positive or negative depending on $\omega$, with the condition that the light shines from $\hat{x}$ or $\hat{y}$ direction. (e) A negative refractive index gives rise to a 3D real image while a positive one gives rise to a virtual image.}
\end{figure}

{\it Angular selectivity.}
Angular selectivity\cite{shen-2014} is easy to achieve in equifrequency Weyl photonic crystals. Assume the Weyl points are at frequency $\omega^*$ and crystal momenta ${\bf k}_i^*=(\pm k^*,0,0)$ or $(0,\pm k^*,0)$, where $i=1,...,4$ label the four Weyl points. If a ray of light in the $xz$ or $yz$ plane with frequency $\omega^*$ has an incident angle $\theta^*=\sin^{-1}(ck^*/\omega^*)$ for which the component of incident momenta parallel to the $xy$ plane ${\bf k}_\parallel={\bf k}_i^*$, it can transmit through the equifrequency PhC; while if ${\bf k}_\parallel\neq{\bf k}_i^*$, it cannot since no available state exists at $\omega^*$, as shown in Fig. \ref{angular1}. Therefore, when a point source of $\omega^*$ shines light against the PhC in its $\hat z$ direction, only four rays transmit partially while others are totally reflected. Note that inside the photonic crystal, the refractive index exactly at frequency $\omega^*$ is not well-defined, and superprism effect occurs\cite{Kosaka-1998}.

{\it Positive and negative refractive index.}
For a PhC with equifrequency Weyl points and nearly isotropic Weyl cones (namely group velocities around the Weyl points are nearly equal), both positive and negative refractive indices can be achieved, as we explain below. For simplicity of illustration, we assume the Weyl cones are isotropic. When a light source is placed above the PhC in its $\hat x$ (or $\hat y$) direction, the momentum in $yz$ (or $xz$) plane is conserved, and the dispersion in this plane is shown in Fig. \ref{cone} \cite{footnote1}. Then it is straightforward to obtain the refractive index: $n(\omega)\equiv\frac{\sin\theta_1}{\sin\theta_2}=\frac{c}{v}(1-\frac{\omega^*}{\omega}) $, which is plotted in Fig. \ref{r} as a function of $\omega$. Remarkably, $n(\omega)$ is positive for $\omega>\omega^*$ and negative for $\omega<\omega^*$.
This property may be used for invisibility cloaking\cite{pendry-2006,leonhardt-2006,schurig-2006}. Moreover, such PhC with a negative refractive index can be used to make superlens\cite{pendry-2000}. In addition, depending on the sign of the refractive index and hence the frequency of the incident light, either 3D real images\cite{notomi-2000} or virtual images are formed, as shown in Fig. \ref{images}. Note that when the Weyl cone is not nearly isotropic, the effective refractive index is not that well-defined as there is no simple proportionality between $\sin\theta_1$ and $\sin\theta_2$. Fortunately, the group velocities of the Weyl cones in the modified DG can be tuned by varying $\epsilon_\textrm{gyr}$, $g_0$ (setting the boundary of the dielectric material), and the ratio of $r/a$ such that it should be feasible to realize nearly isotropic Weyl cones.

\begin{figure}
    \subfigure[]{\includegraphics[width=3.8cm]{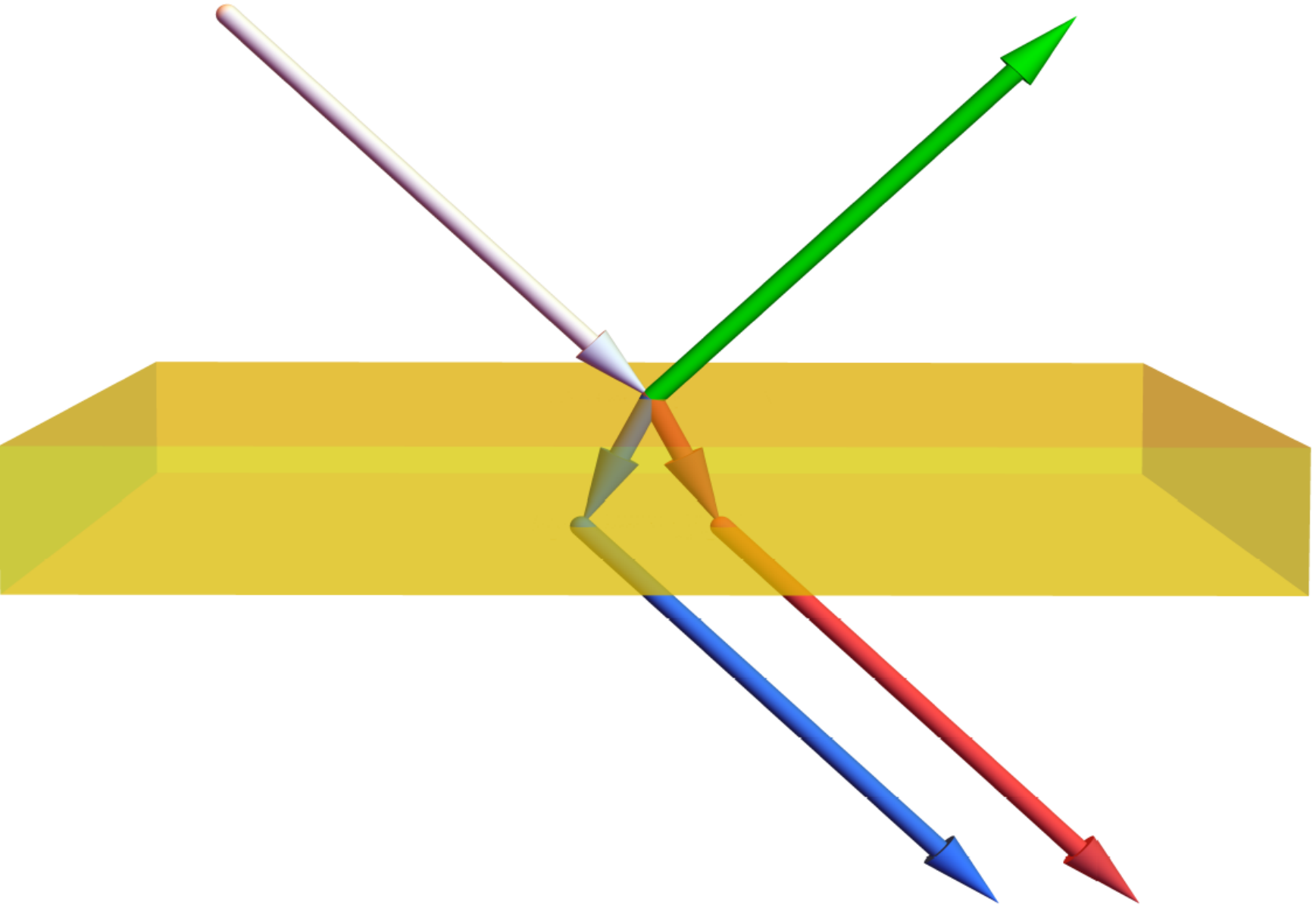}\label{frequency1}}~~~~
	\subfigure[]{\includegraphics[width=3.8cm]{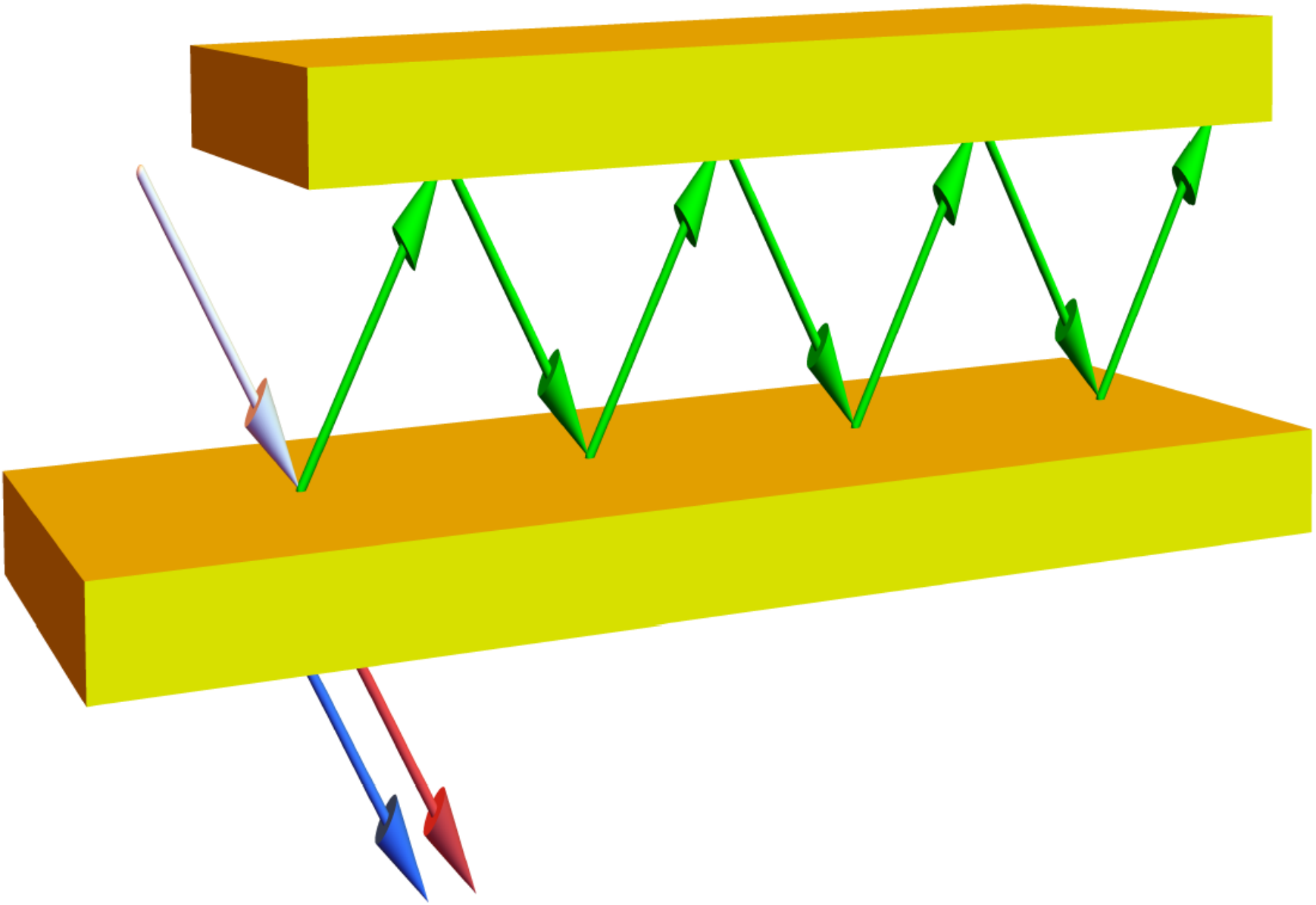}\label{frequency2}}
	\caption{(a) A collimated beam with a broad range of frequencies incident at an angle $\theta$, close to $\theta^*$. The ``green'' ray in the frequency range $(\omega^*-\delta\omega,\omega^*+\delta\omega)$ can be totally reflected while lights with other frequencies, represented by ``blue'' and ``red'' rays, can transmit.
(b) Based on this novel frequency selectivity, a waveguide of light with a particular range of frequencies can be made from the photonic crystal with equifrequency Weyl points.}
\end{figure}

{\it Frequency selectivity}. Another advantage of a PhC with equifrequency Weyl points is high-quality frequency-selectivity. Assume a collimated light with a broad range of frequencies, including $\omega^*$, is incident onto the PhC in the $xz$ plane, as shown in Fig. \ref{frequency1}.
If the incident angle $\theta=\theta^*+\delta \theta$ with $\delta\theta\neq 0$, the light with frequency in the range $(\omega^*-\delta\omega,\omega^*+\delta\omega)$ will be totally reflected, where $\delta\omega\approx (v/c)\cos(\theta^*)\omega^*|\delta\theta|$ assuming $v/c\ll 1$ and $\delta\theta\ll 1$, while the light with other frequencies can be partially transmitted into the PhC. This novel property is illustrated in Fig. \ref{frequency1}, where the ``green'' ray represents light in the range of $(\omega^*-\delta\omega,\omega^*+\delta\omega)$ and thus is totally reflected, while the ``blue'' and ``red'' rays represent lights with either higher or lower frequencies that can be partially transmitted into the PhC. By adjusting the incident angle $\theta$, close to $\theta^*$, we can control the frequency range of total reflections. This controllable frequency-selectivity is of practical importance.
For instance, mirrors, filters, and waveguides (see Fig. \ref{frequency2}) of a particular range of frequencies can be made of photonic crystals with equifrequency Weyl points. In Fig. \ref{frequency2}, the partially reflected ``blue'' and ``red'' rays decay to negligible amount after a few times of reflections, while only the totally reflected ``green'' ray is trapped in the waveguide.

{\it Conclusion and discussions}. We have shown that a minimal number of four equifrequency Weyl points can be realized in the modified time-reversal-invariant DG PhC by both general symmetry analysis and the first-principle calculations of a particular modified DG PhC with symmetry $D_{2d}$.
Our proposal of introducing four air spheres into each primitive cell of the DG PhC is experimentally feasible.
Such topological PhC with a minimal number of four equifrequency Weyl points may pave the way to future applications including angular selectivity, frequency selectivity, invisibility cloaking, 3D-imaging, and large volume single mode lasers\cite{chua-2014} and to novel topological phenomena such as topological Imbert-Fedorov shift\cite{onoda-2004,bliokh-2008,xie-2015}.

{\it Acknowledgement}. This work was supported by the National Thousand-Young-Talents Program and the NSFC under Grant No. 11474175 at Tsinghua University.

\appendix
\section{The effective $k\!\cdot\! p$ theory for modified double-gyroid and equifrequency Weyl points}

In this Appendix, we give an explicit description of the effective $k\!\cdot\! p$ theory of modified double-gyroid (DG) photonic crystals.  The low energy theory near the triple degeneracy at the $\Gamma$ point for unmodified DG PhC is given by Eq. (1) in the main text, where the 3$\times$3 matrices are explicitly given in terms of $(L_i)_{jk}=i \epsilon_{ijk}$. From $L_i$ matrices, one can construct all independent 3$\times$3 Hermitian matrices\cite{luttinger-1956}: $1_{3\times 3}$; $(L_x, L_y, L_z)$; $(\{L_x, L_y\}, \{L_y, L_z\}, \{L_z, L_x\})$ and $(L_z^2, L_x^2-L_y^2)$, forming A, T$_1$, T$_2$ and E representation of the $O_h$ group, respectively.

As stated in the main text, we consider the perturbations lowering $O_h$ symmetry to $D_{2d}$ symmetry.
The most relevant perturbation near the $\Gamma$ point is a constant term which does not depend on $k$. As there is only one such term (proportional to $L^2_z$), which respects
the $D_{2d}$ symmetry. Thus we have $H_1$. The aforementioned line node in the $k_z=0$ plane can be obtained by diagonalizing $H_0+H_1$ directly. This line node is protected by the $\sigma_h$ symmetry. To illustrate, the dispersions along the $k_x$ axis are given by: $E_{1a}=\al_1 k_x^2+ \beta_1$, $E_{1b}=(\al_1+\al_2) k_x^2+ \beta_1$ and $E_2=(\al_1+\al_2) k_x^2-2\beta_1$, where the subscripts 1 and 2 denote the reflection subspaces while $a,b$ denote the states in one of the reflection subspace. Among the range that $\al_2 \beta_1>0$, the crossing of two bands is inevitable, namely, the crossing between $E_{1a}$ and $E_2$ must occur. The crossing along $k_x$ axis is given by $k_x=\pm \sqrt{3\beta_1/\al_2}$. Since they are in different subspaces, the hybridization is forbidden so long as the reflection symmetry is present.

The next-relevant terms are those that are linear in momentum. In DG PhC, the time-reversal symmetry is present without applying external magnetic field. Since momentum is odd under the time reversal operation T, only the time reversal odd matrices, i.e., $L_i$ have to be taken into considerations at linear level. The T invariant perturbation is given by $H_2(\vec k)=\sum_{ij} g_{ij} k_i L_j$ for a real matrix $g$. Restricting to $D_{2d}$ symmetry, only two elements of $g$ survive, namely, $g_{11}=-g_{22} \equiv \beta_2$. This explains $H_2(\vec{k})$ in the main text. In the presence of perturbations represented by $H_2(\vec{k})$, the line node splits except at four discrete points at the $k_x$ and $k_y$ axes. We will explicitly show that these gapless points are equifrequency Weyl points below.

We find that the two bands cross in $k_x$ and $k_y$ axes in the presence of $H_1$ within a broad range of parameters. These gapless points survive as long as the perturbations respect $C_{2T}$ symmetry that mentioned in the main text. For illustration, we consider the $C_2$ symmetry along the $y$ axis (the space group is non-symmorphic, which is not important here). The $C_{2T}$ is given by $C_{2T}= e^{-i \pi L_y} e^{i \pi L_y} K= K$, where $K$ denotes the complex conjugation. Thus the Hamiltonian in the $k_y=0$ plane is real. For such a two-dimensional subsystem, a topological invariant can be defined by the first homotopy group of projector\cite{fang-2015}, $\pi_1(\frac{O(M+N)}{O(M)\oplus O(N)})=Z_2$.  For a gapless point (like those in $k_x$ axis as we consider) enclosed by a topological nontrivial invariant line in such two-dimensional system, it will not disappear from $k_y=0$ plane unless it annihilates with another gapless point. There is another $C_2$ symmetry along the $z$ axis, so the gapless points will be pinned in the $k_x$ axis. For the same reason, the $k_y$ axis also possesses two gapless points. This is consistent with $S_4$ symmetry in $D_{2d}$ group.

In order to show the Weyl points explicitly, we first find the positions of gapless points. A straightforward calculation leads to four positions, two at the $k_x$ axis, $(\pm k^\ast,0,0)$, and two at the $k_y$ axis, $(0, \pm k^\ast,0)$, where $k^\ast=\Delta/\al_2$ with $\Delta=\sqrt{3\al_2 \beta_1+ \beta_2^2}$. It is crucial that $\Delta>0$ for the existence of gapless points. Note that this is equivalent to $ \al_2 \beta_1>0$, consistent with the analysis of line nodes above. Next, we transform the original basis into the one that diagonalizes $L_x$ and then project the Hamiltonian into the two bands that cross at the gapless points. After a tedious but straightforward calculation, the explicit form of Hamiltonian at $(k^\ast,0,0)$ is given by Eq. (2) in the main text, with
\bea
	 v_x'&=& \frac{\Delta}{\Delta^2+\beta_2^2}[(2/\delta+1)\Delta^2+2\beta_2^2 \delta],\\
	 v_x&=&-\frac{\Delta^{3}}{\Delta^2+\beta_2^2},\\
     v_y&=& - (1+\frac{\al_3}{\al_2})\times \frac{\Delta \beta_2}{\sqrt{\Delta^2+\beta_2^2}},\\
	 v_z&=& -\frac{\al_3}{\al_2} \times \frac{\Delta^2}{\sqrt{\Delta^2+\beta_2^2}},
\eea
where $\delta=\al_2/\al_1$. It is apparent that the gapless point is a Weyl point and its chirality is given by the sign of $\chi=-\beta_2 \frac{\al_3}{\al_2}(1+\frac{\al_3}{\al_2})$.


\begin{thebibliography}{99}
\bibitem{mpb} The calculations in this paper were performed using the MIT Photonic-Bands (MPB) package.
\bibitem{footnote1} When the light source with frequency $\omega$ is put above the PhC in the $\hat {z}$ direction, an effective refractive index satisfying Snell's law cannot be well-defined.
\end{thebibliography}
\end{document}